\documentclass[10pt,a4paper]{article}
\usepackage{amsmath}
\usepackage{amsfonts}
\usepackage{amssymb}
\usepackage{graphicx}
\usepackage{textcomp}
\usepackage{url}
\usepackage{lipsum}
\usepackage{setspace}
\usepackage{comment}
\usepackage{color}
\usepackage[numbers,sort&compress]{natbib}
\usepackage[usenames,dvipsnames,svgnames,table]{xcolor}
\usepackage[lofdepth,lotdepth,caption=false]{subfig}
\usepackage[pdfstartview=FitH]{hyperref}
\usepackage{hyperref}
\newsavebox{\bigleftbox}
\hypersetup{
 colorlinks,
 citecolor=Blue,
 linkcolor=Blue,
 urlcolor=Blue
}

\makeatletter
 \def\footnoterule{\kern-3\p@
   \noindent\hrulefill \kern 2.8\p@} 
\makeatother
\usepackage[left=3.00cm, right=3.00cm, top=2.00cm, bottom=2.00cm]{geometry}

\title{\textbf{Exploring the Electronic and Mechanical Properties of the Recently
Synthesized Nitrogen-Doped Monolayer Amorphous Carbon}}
\author{
   E. J. A. dos Santos$^{1,2}$, M. L. Pereira Junior$^{3}$, R. M. Tromer$^{4}$, D. S. Galvão$^{5}$, \\ and L. A. Ribeiro Junior$^{1,2,\S}$
	}
\date{}

\begin{document}
    \maketitle
	\vspace{-0.6cm}
	\begin{center}\small
    \textit{$^{1}$Institute of Physics, University of Bras\'ilia, 70910-900, Bras\'ilia, Brazil}\\
	\textit{$^{2}$Computational Materials Laboratory, LCCMat, Institute of Physics, University of Bras\'ilia, 70910-900, Bras\'ilia, Brazil}\\
    \textit{$^{3}$ University of Bras\'{i}lia, College of Technology, Department of Electrical Engineering, Bras\'{i}lia, Federal District, Brazil}\\
    \textit{$^{4}$ School of Engineering, MackGraphe, Mackenzie Presbyterian University, São Paulo, São Paulo, Brazil}\\
    \textit{$^{5}$ Department of Applied Physics and Center for Computational Engineering and Sciences, State University of Campinas, Campinas, São Paulo, Brazil} \\
    \phantom{.}\\ \hfill
        $^{\S}$\url{ribeirojr@unb.br}\hfill
		\phantom{.}
	\end{center}
	

\onehalfspace 

\noindent\textbf{Abstract: The recent synthesis of nitrogen-doped monolayer amorphous carbon (MAC @N) opens new possibilities for multifunctional materials. In this study, we have investigated the nitrogen doping limits and their effects on MAC@N's structural and electronic properties using density functional-based tight-binding simulations. Our results show that MAC@N remains stable up to 35\% nitrogen doping, beyond which the lattice becomes unstable. The formation energies of MAC@N are higher than those of nitrogen-doped graphene for all the cases we have investigated. Both undoped MAC and MAC@N exhibit metallic behavior, although only MAC features a Dirac-like cone. MAC has an estimated Young's modulus value of about 410 GPa, while MAC@N's modulus can vary around 416 GPa depending on nitrogen content. MAC displays optical activity in the ultraviolet range, whereas MAC@N features light absorption within the infrared and visible ranges, suggesting potential for distinct optoelectronic applications. Their structural thermal stabilities were addressed through molecular dynamics simulations. MAC melts at approximately 4900K, while MAC@N loses its structural integrity for temperatures ranging from 300K to 3300K, lower than graphene. These results point to potential MAC@N applications in flexible electronics and optoelectronics.}

\section{Introduction}

Two-dimensional (2D) carbon-based materials \cite{hirsch2010era}, particularly graphene \cite{geim2009graphene}, have received significant attention over the past decades due to their exceptional mechanical, electronic, and optical properties \cite{jana2021emerging}. Graphene, a monolayer of sp$^{2}$-hybridized carbon atoms arranged in a honeycomb lattice, has become a cornerstone for developing flexible electronics, energy storage devices, and other applications \cite{kumar2018recent}. The discovery of graphene has renewed the search for other 2D carbon allotropes, aiming to explore new physical properties and functionalities by modifying the carbon lattice \cite{enyashin2011graphene,fan2021biphenylene,hou2022synthesis,meirzadeh2023few}. Among these new materials, pure \cite{toh2020synthesis,tian2023disorder} and doped \cite{bai2024nitrogen,zhang2022structure} amorphous 2D carbon lattices have emerged as promising candidates due to their inherent structural flexibility and tunable electronic characteristics.

The first freestanding monolayer of amorphous carbon (MAC) \cite{toh2020synthesis} was recently synthesized using laser-assisted chemical vapor deposition, marking a significant breakthrough in 2D materials. MAC comprises a mixture of sp$^{2}$ and sp$^{3}$ -like hybridized carbon atoms randomly arranged in five- to eight-member rings. Unlike the crystalline structure of graphene, MAC exhibits short-range order with a wide distribution of bond lengths and angles, significantly impacting its mechanical and electronic properties. This unique ring distribution, distinct from traditional Zachariasen random networks \cite{joo2017realization}, results in a structure with remarkable thermal stability and mechanical strength \cite{felix2020mechanical,junior2021reactive}. As a result, MAC holds potential for applications in flexible electronics \cite{gastellu2022electronic}, energy storage \cite{xia2020monolayer}, and optoelectronics \cite{garzon2022optoelectronic}, similar to graphene \cite{avouris2012graphene} but with the added benefits from the structural disorder.

The quite recent synthesis of nitrogen-doped monolayer amorphous carbon (MAC@N) introduces another exciting development in two-dimensional materials \cite{bai2024nitrogen}. MAC@N was synthesized through a space-confined solution-phase process using a layered double hydroxide (LDH) template, facilitating precise control over nitrogen incorporation into the amorphous carbon lattice. This novel approach yields a freestanding monolayer consisting of a random arrangement of five-, six-, and seven-membered carbon rings, where nitrogen atoms replace some of the carbon atoms within the network. The polymerization of pyrrole within the confined interlayer of the LDH template is critical to forming the nitrogen-doped structure, preventing excessive bond rearrangements that could lead to instability during synthesis. As a result, MAC@N exhibits a unique in-plane $\pi$-like-conjugation electronic structure, setting it apart from other carbon-based 2D materials. Nitrogen atoms further impact the material’s electronic and mechanical properties by introducing localized electronic states distinct from those observed in pure amorphous carbon or crystalline graphene \cite{bai2024nitrogen}. Additionally, the nitrogen-doped structure maintains the flexibility and structural disorder characteristic of amorphous carbon while offering enhanced tunability of its electronic and optical properties through controlling nitrogen content \cite{bai2024nitrogen}. These features make MAC@N a promising candidate for applications in flexible electronics, optoelectronics, and other multifunctional devices, where structural flexibility and controlled electronic behavior are critical.

Nitrogen doping is a well-known strategy for tuning the properties of carbon-based materials \cite{mortazavi2019prediction,mortazavi2019outstanding,mortazavi2018n,mortazavi2023electronic}, providing enhanced control over electronic behavior \cite{mortazavi2020nanoporous}, mechanical strength \cite{mortazavi2021ultrahigh,mortazavi2022anisotropic}, and chemical reactivity \cite{wu2016tuning,talukder2021nitrogen}. However, the critical issue of the nitrogen saturation limit that can be incorporated into the amorphous carbon lattice while maintaining its structural integrity remains to be determined. Understanding this doping limit is crucial for optimizing MAC@N sheets for various applications, particularly in flexible electronics and optoelectronics, where precise control over material properties is essential.

In this study, we have addressed this fundamental question through density functional-based tight-binding (DFTB+) simulations. Molecular dynamics (MD) simulations were also carried out to investigate the nitrogen doping limits in MAC@N and their effects on the material's structural, optical, mechanical, and electronic properties. Our findings revealed that MAC@N remains structurally stable up to a nitrogen doping concentration of 35\%, beyond which the lattice becomes unstable. Furthermore, we have also analyzed these properties for nitrogen-doped graphene (graphene@N) and MAC@N to understand how nitrogen incorporation comprehensively alters their behavior. 

\section{Methodology}

In this study, we have used DFTB+ simulations \cite{hourahine2020dftb+,aradi2007dftb+} to explore the structural, electronic, and mechanical properties of MAC and MAC@N  with different nitrogen-doping contents. The 3ob set of Slater-Koster parameters \cite{gaus2013parametrization,gaus2014parameterization}, and dispersion correction were adopted for the calculations. Importantly, although providing lower precision results than pure DFT methods, DFTB+ can produce good-quality results for treating large-scale systems, which would be cost-prohibitive with pure DFT methods. This is particularly important for modeling the disordered structure of monolayer amorphous carbon, which typically requires large unit cells.

\begin{figure}[!htb]
    \centering
    \includegraphics[width=\linewidth]{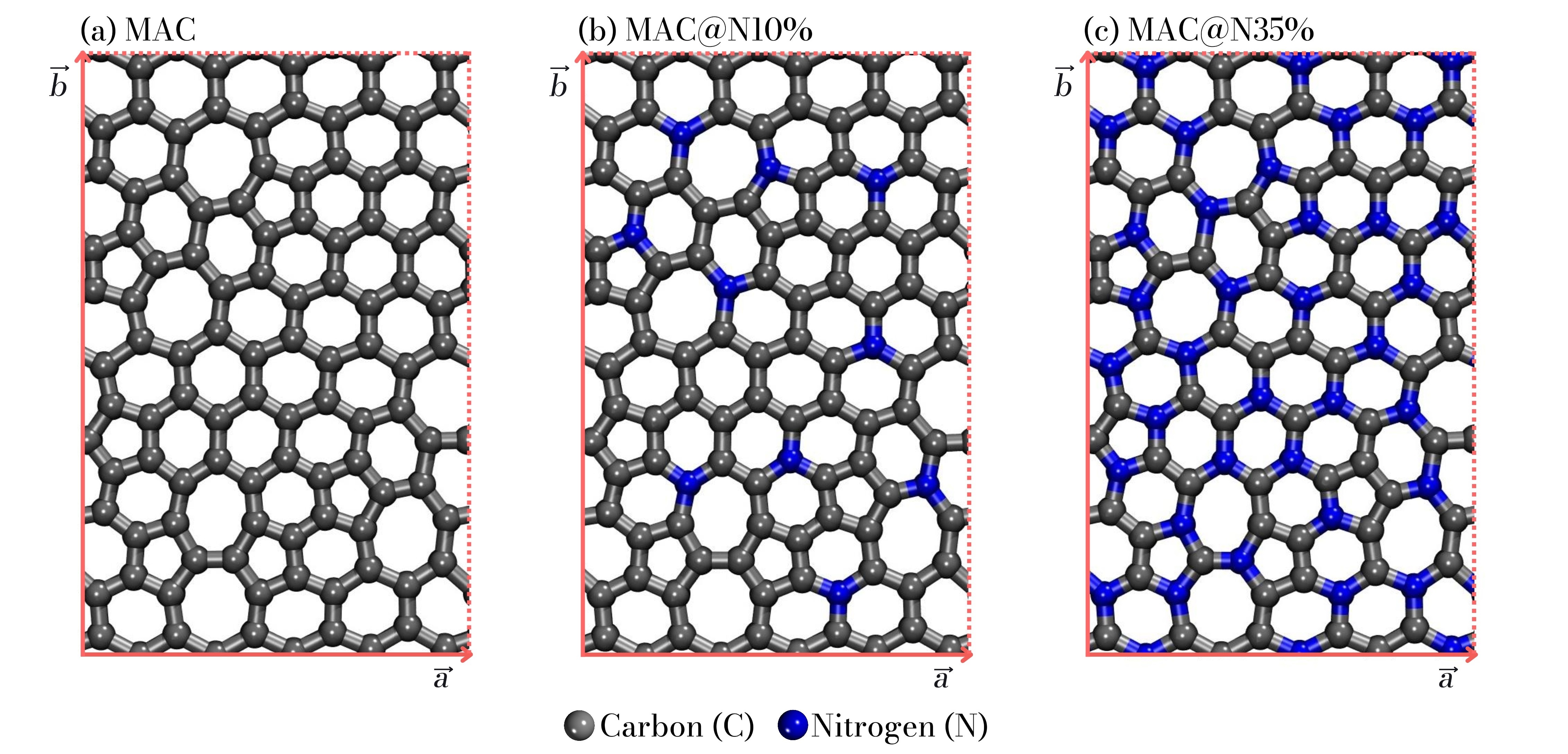}
    \caption{Schematic representation of some of the proposed structural models with their corresponding lattice vectors. (a) Pure amorphous monolayer carbon (MAC), (b) MAC@N structure doped with 10\% nitrogen, and (c) MAC@N structure doped with 35\% nitrogen.}
    \label{fig:system}
\end{figure}

Typical MAC, graphene@N, and MAC@N structural models (P1-CS-1 space group) contain dozens of atoms and were created based on the available experimental data \cite{bai2024nitrogen} (see Figure \ref{fig:system}). The lattice vectors $\vec{a}$ and $\vec{b}$ are 15.00 \r{A} and 20.42 \r{A}, respectively. These models consist of randomly distributed five-, six-, and seven-membered carbon rings and nitrogen-doping sites, forming the MAC and MAC@N sheets. The nitrogen doping process was simulated by randomly replacing carbon atoms with nitrogen ones at different concentrations, ranging from 5\% up to 40\% of the total number of carbon atoms. Figure \ref{fig:system} illustrates the model lattices studied in this work, i.e., the (a) MAC, (b) MAC@N structure doped with 10\% nitrogen (MAC@N-10\%), and (c) MAC@N structure doped with 35\% nitrogen (MAC@N-35\%).

The geometry optimizations were performed using the DFTB+ package for each nitrogen-doped structure. The convergence criteria for structure optimization and energy calculation were set to energy tolerance of 0.02 kcal/mol, maximum force tolerance of 0.1 kcal/mol/\r{A}, and maximum displacement tolerance of 0.001 \r{A}, ensuring that each system was in its lowest energy state. Periodic boundary conditions were applied to mimic the behavior of an infinite monolayer. The optimized geometries were then used to analyze the electronic and mechanical properties of the MAC and MAC@N systems.

The electronic and optical properties were also computed, including the electronic band structure, the projected density of states (DOS), charge distribution, and light absorption coefficient. The Monkhorst-Pack scheme was employed to sample the Brillouin zone with a $15\times15\times1$ k-point grid for accurate results. The mechanical properties of the nitrogen-doped monolayer amorphous carbon were assessed by calculating Young’s modulus and Poisson’s ratio for each optimized structure. Special attention was dedicated to the effects of nitrogen doping on the electronic characteristics of the MAC@N systems, particularly on the bandgap values and the presence or absence of Dirac-like cones. 

To further explore the thermal stability of MAC and MAC@N layers, we also carried out MD simulations using DFTB+. These simulations were performed under the NVT ensemble using a time step of 1 fs and a Nosé-Hoover thermostat \cite{shuichi1991constant} to control the temperature values. The systems were incrementally heated from 10K to 5000K to determine their melting points. 

\section{Results and Discussions}

\begin{figure}[!htb]
    \centering
    \includegraphics[width=\linewidth]{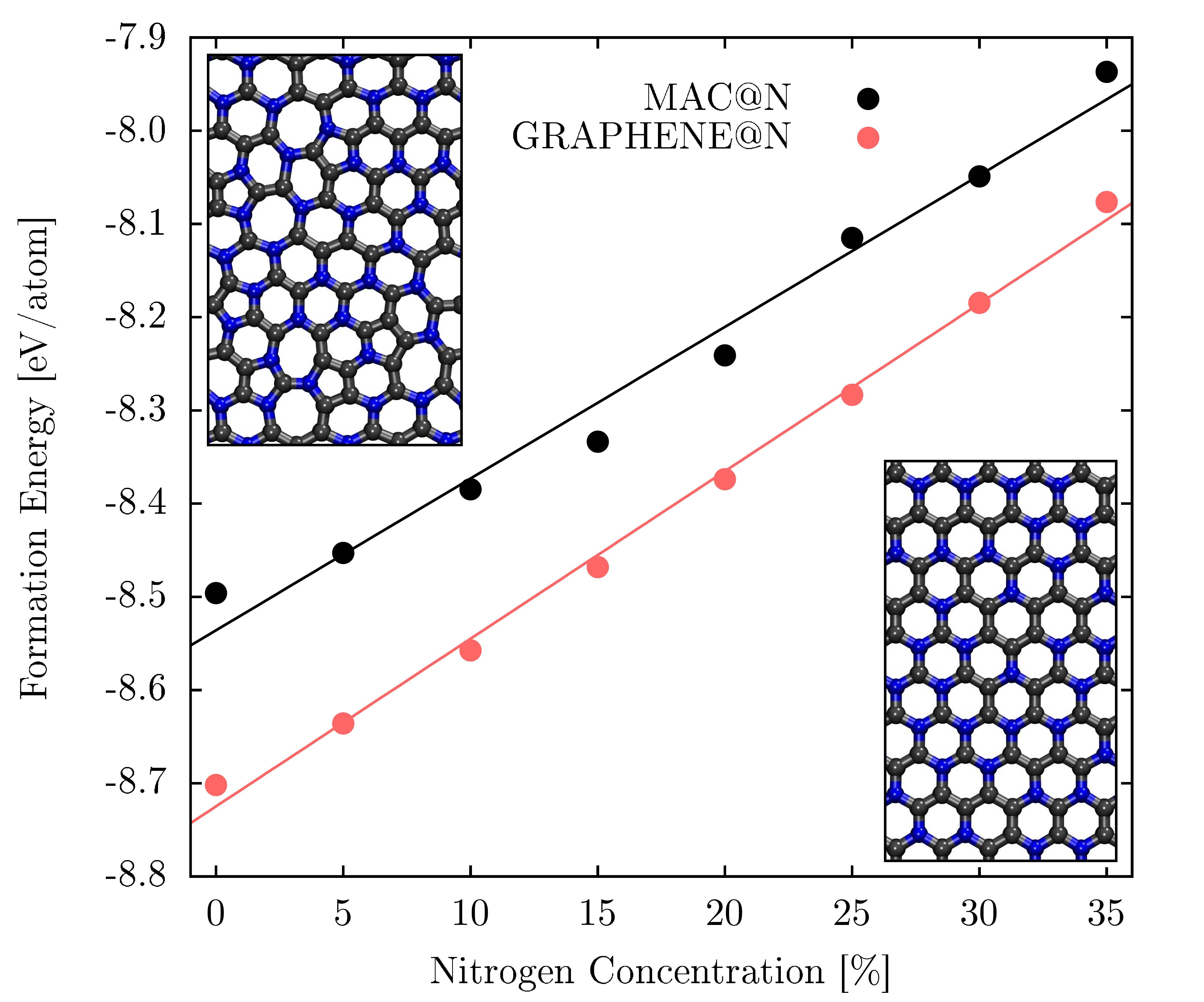}
    \caption{Formation energy values as a function of nitrogen concentration (N\%) for MAC@N (black) and graphene@N (red). The insets depict the atomic structures of the MAC@N (left) and graphene@N (right) systems, with the nitrogen atoms highlighted in blue.}
    \label{fig:energy}
\end{figure}

We begin the discussions by analyzing the structural results for graphene@N and MAC@N. As shown in Figure \ref{fig:energy}, the formation energy of MAC@N is consistently higher than that of graphene@N for all nitrogen contents investigated here. This difference arises from the intrinsic structural characteristics of the two materials. Graphene possesses a highly ordered hexagonal lattice, allowing for uniform incorporation of nitrogen atoms with minimal structural disruption to the bonding network. In contrast, MAC exhibits a disordered structure composed of a mixture of five-, six-, and seven-membered carbon rings, which leads to variations in bond lengths and angles. When nitrogen atoms are introduced into the disordered MAC lattice, they disrupt the local bonding environment more significantly than in the regular structure of graphene. 

Also, nitrogen atoms have different atomic radii and bonding configurations than carbon ones. As the lattice has to accommodate more nitrogen atoms at higher doping concentrations, it results in more pronounced bond distortions and increased residual strain throughout the structure. This additional strain increases the formation energy values as the lattice becomes progressively less stable with increasing nitrogen content.

Although for graphene@ and MAC@N, the increase in formation energy values allows a good linear fitting with nitrogen concentration, their behavior is slightly different. For graphene@N, the fitting is almost perfectly linear, reflecting the uniform nature of its hexagonal carbon lattice. The ordered structure of graphene allows for an easier adjustment to nitrogen incorporation, resulting in a steady increase in formation energy with each incremental increase in doping concentration. In contrast, MAC@N shows a slight deviation from ideal linearity, with some dispersion among data points. This deviation arises from the amorphous nature of the MAC lattice, where variations in local atomic environments lead to uneven strain distributions as nitrogen atoms are incorporated. The intrinsic disorder of MAC introduces variations in how nitrogen atoms interact with the surrounding carbon network, resulting in differences in the energy values required to stabilize the structure at different doping levels.

The results further reveal that the nitrogen doping limit for maintaining lattice structural stability in MAC@N is 35\%. In contrast, graphene can accommodate a higher nitrogen concentration before becoming unstable \cite{shi2015much}. This difference is primarily attributed to the structural characteristics of MAC, which, unlike graphene, lacks uniformity and long-range order \cite{toh2020synthesis,bai2024nitrogen,tian2023disorder}. The disordered arrangement of carbon rings in MAC results in a structure that experiences significant structural deformations at elevated doping levels. As discussed above, as more nitrogen atoms are incorporated into the structures, there is an increase in strain, which contributes to significant structural distortions. In contrast, the regular hexagonal lattice of graphene offers a more stable framework for nitrogen incorporation, enabling the material to sustain higher doping levels without compromising its structural integrity.

Based on the formation energy and structural analyses, we selected two relevant nitrogen doping concentrations for a more detailed investigation: 10\% and 35\%, respectively. The 10\% doping level was chosen to align with the concentration reported in the experimental synthesis of MAC@N \cite{bai2024nitrogen}. This choice allows for a direct comparison between our theoretical predictions and experimental observations, providing a deeper understanding of MAC@N's properties at a moderate doping concentration. In contrast, the 35\% doping level was selected because it represents the upper limit of nitrogen concentration at which the MAC@N lattice remains stable, offering insights into the structural and electronic behavior of the material at the highest achievable doping level without compromising lattice structural stability.

\begin{figure}[!htb]
    \centering
    \includegraphics[width=\linewidth]{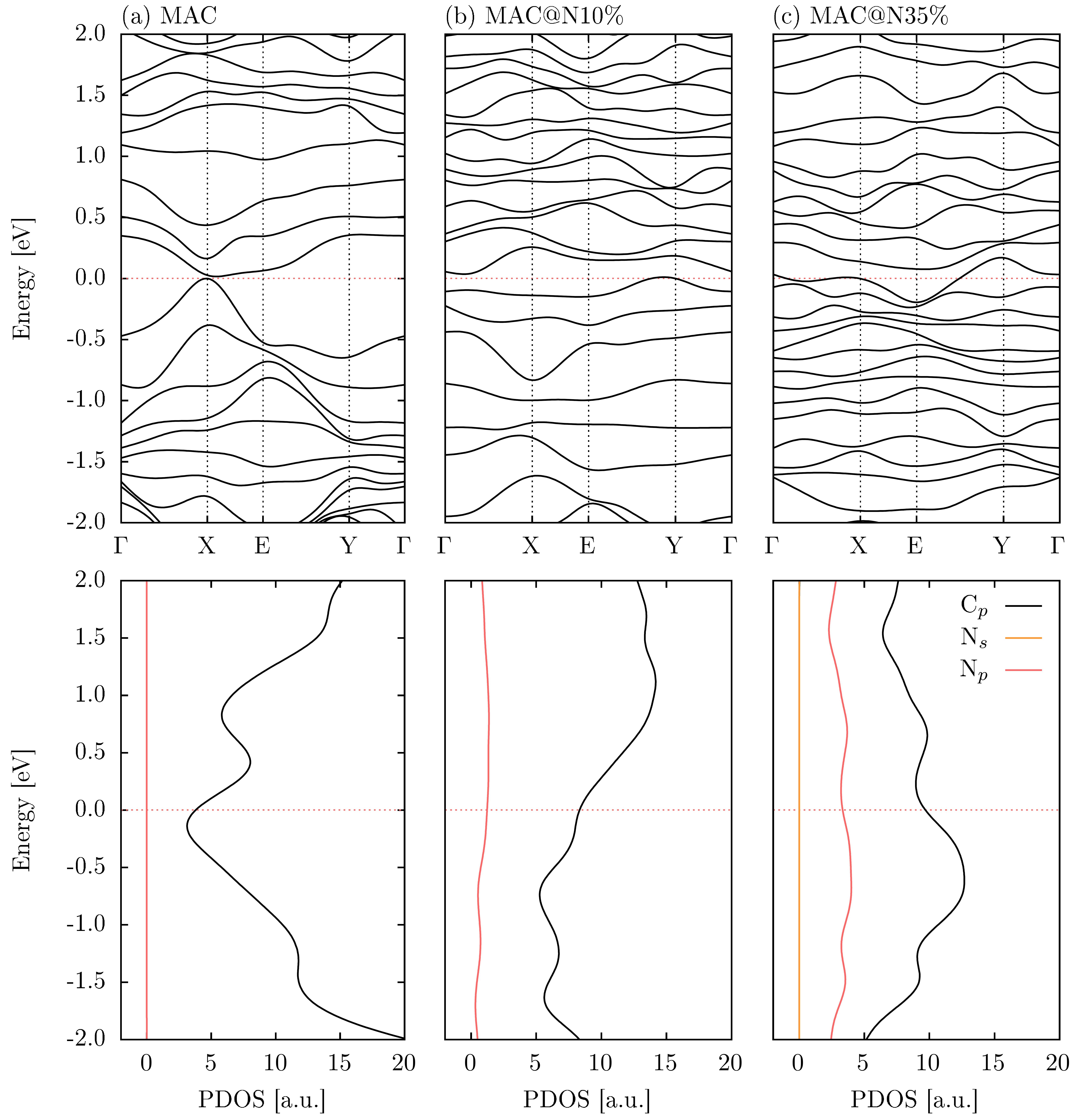}
    \caption{Electronic band structure and projected density of states (PDOS) for (a) MAC, (b) MAC@N-10\%, and (c) MAC@N-35\%. The red dashed line indicates the Fermi level.}
    \label{fig:bandas-dos}
\end{figure}

The electronic band structure features change as nitrogen atoms are incorporated into the MAC lattice. For the MAC@N10\% case (Figure \ref{fig:bandas-dos}(b)), the Dirac-like cone observed in undoped MAC disappears. This change can be attributed to the delocalized $\pi$-electron network disruption caused by nitrogen substitution. Due to their distinct number of electrons and bond nature compared to carbon, nitrogen atoms introduce localized electronic states into the band structure. These localized states disrupt the symmetry that initially enabled the appearance of the Dirac-like cone in the undoped MAC. The PDOS for MAC@N with 10\% nitrogen shows that carbon p-orbitals dominate the contributions to electronic states near the Fermi level. In contrast, the nitrogen p-orbitals contribute only marginally. This trend indicates that, at this doping level, nitrogen primarily acts as a perturbative element in the electronic structure rather than playing a significant role in the material's conduction mechanism.

In contrast, the electronic band structure of the MAC@N system with 35\% nitrogen doping (Figure \ref{fig:bandas-dos}(d)) exhibits more pronounced changes. The PDOS for this system reveals relevant contributions from nitrogen's s- and p-orbitals to the electronic states near the Fermi level. This behavior differs from the 10\% nitrogen-doped system, where the nitrogen p-orbitals played a more limited role. The increased nitrogen content at 35\% leads to a more significant localization of electronic states, particularly those arising from nitrogen's s-orbitals, which hybridize with the carbon p-orbitals to form the bands near the Fermi level. This enhanced hybridization alters the electronic structure, resulting in a distinct band profile compared to the undoped MAC and MAC@N10\% systems.

\begin{figure}
    \centering
    \includegraphics[width=\linewidth]{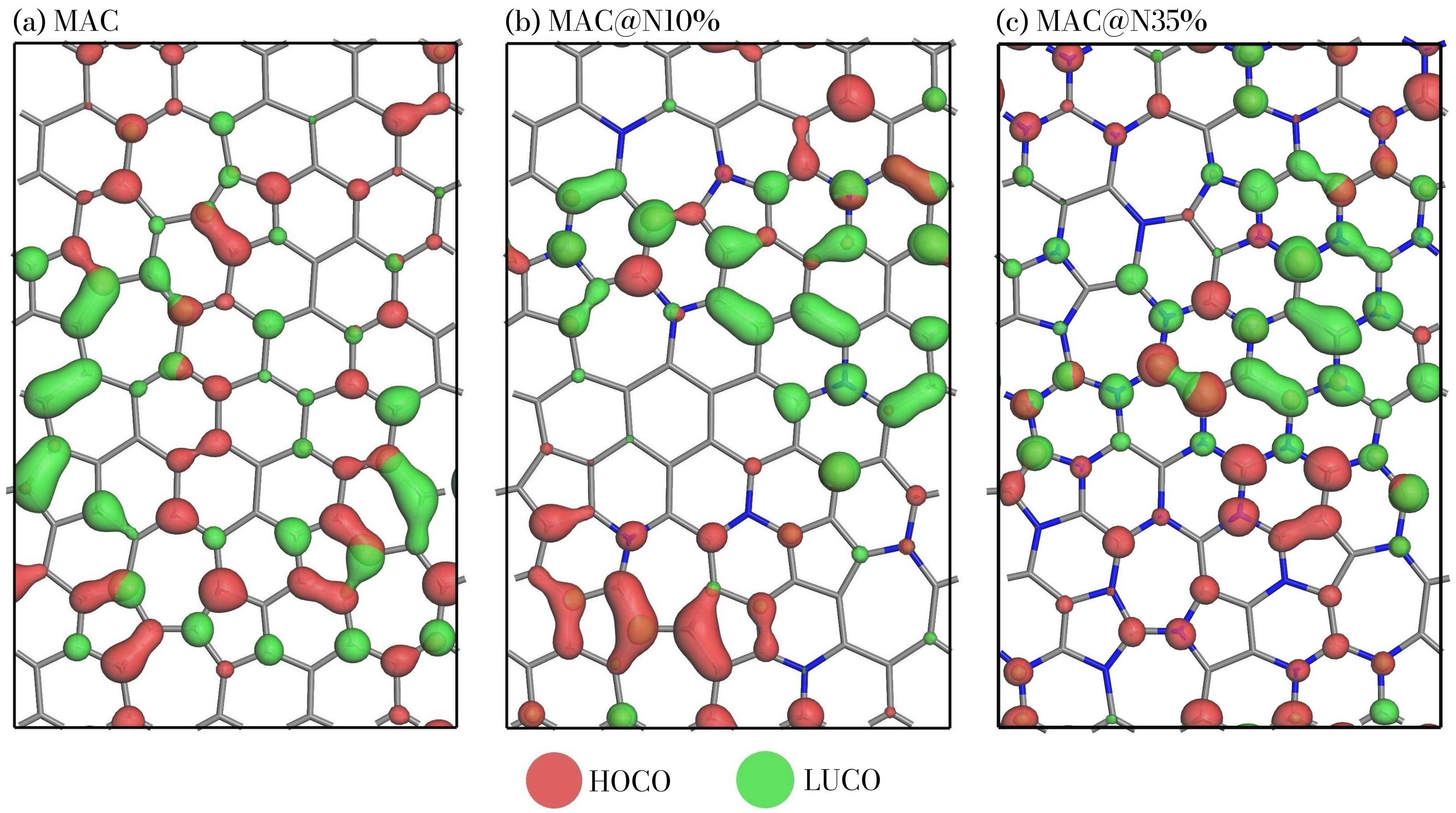}
    \caption{Spatial distribution of the Highest Occupied Crystalline Orbital (HOCO, in red) and Lowest Unoccupied Crystalline Orbital (LUCO, in green) for (a) MAC, (b) MAC@N-10\%, and (c) MAC@N-35\%.}
    \label{fig:orbital}
\end{figure}

Figure \ref{fig:orbital} illustrates the spatial distribution of the Highest Occupied Crystalline Orbital (HOCO, in red) and Lowest Unoccupied Crystalline Orbital (LUCO, in green) for the three systems: undoped MAC (Fig. \ref{fig:orbital}(a)), MAC@N-10\% (Fig. \ref{fig:orbital}(b)), and MAC@N-35\% (Fig. \ref{fig:orbital}(c)). Due to the amorphous nature of these materials, the overall spatial distribution of the HOCO and LUCO lacks a clear or ordered pattern, with distinct contributions from atoms at different parts of the structure. This is characteristic of amorphous materials, where the lack of long-range atomic order leads to localized electronic states \cite{bai2024nitrogen}. The observed orbital delocalization in undoped MAC aligns closely with findings from previous studies \cite{tromer2024structural,zhang2022structure,tromer2021optoelectronic}.

Introducing nitrogen into the MAC structure does not fundamentally alter the spatial distribution of frontier electronic orbitals. However, as shown in Figure \ref{fig:orbital}(b), regions with more nitrogen atoms tend to exhibit more localized orbitals. This behavior can be attributed to the electronic nature of nitrogen, which is more electronegative than carbon. In areas where nitrogen atoms cluster or are more concentrated, they create localized perturbations in the electronic structure, leading to the confinement of the HOCO and LUCO in those regions. Consequently, these orbitals become more localized around nitrogen-rich areas. Nonetheless, the overall spatial distribution of the orbitals persists due to the amorphous nature of the MAC structure.

At higher nitrogen concentrations, as observed in the MAC@N system with 35\% doping (Figure \ref{fig:orbital}(c)), there is an apparent increase in the degree of orbital delocalization. This trend can be explained by the more significant number of nitrogen-induced electronic states interacting with the carbon network. As more nitrogen atoms are incorporated, they create more hybridized states with carbon p-orbitals, resulting in a broader overlap of electronic states. This enhanced hybridization between nitrogen and carbon atomic orbitals increases the coupling of the electronic orbitals throughout the structure, leading to greater delocalization of both the HOCO and LUCO. The more extensive overlap between nitrogen and carbon orbitals allows the electronic states to spread further across the structure, reducing the tendency for orbitals to remain confined to localized regions.

\begin{figure}[!htb]
    \centering
    \includegraphics[width=\linewidth]{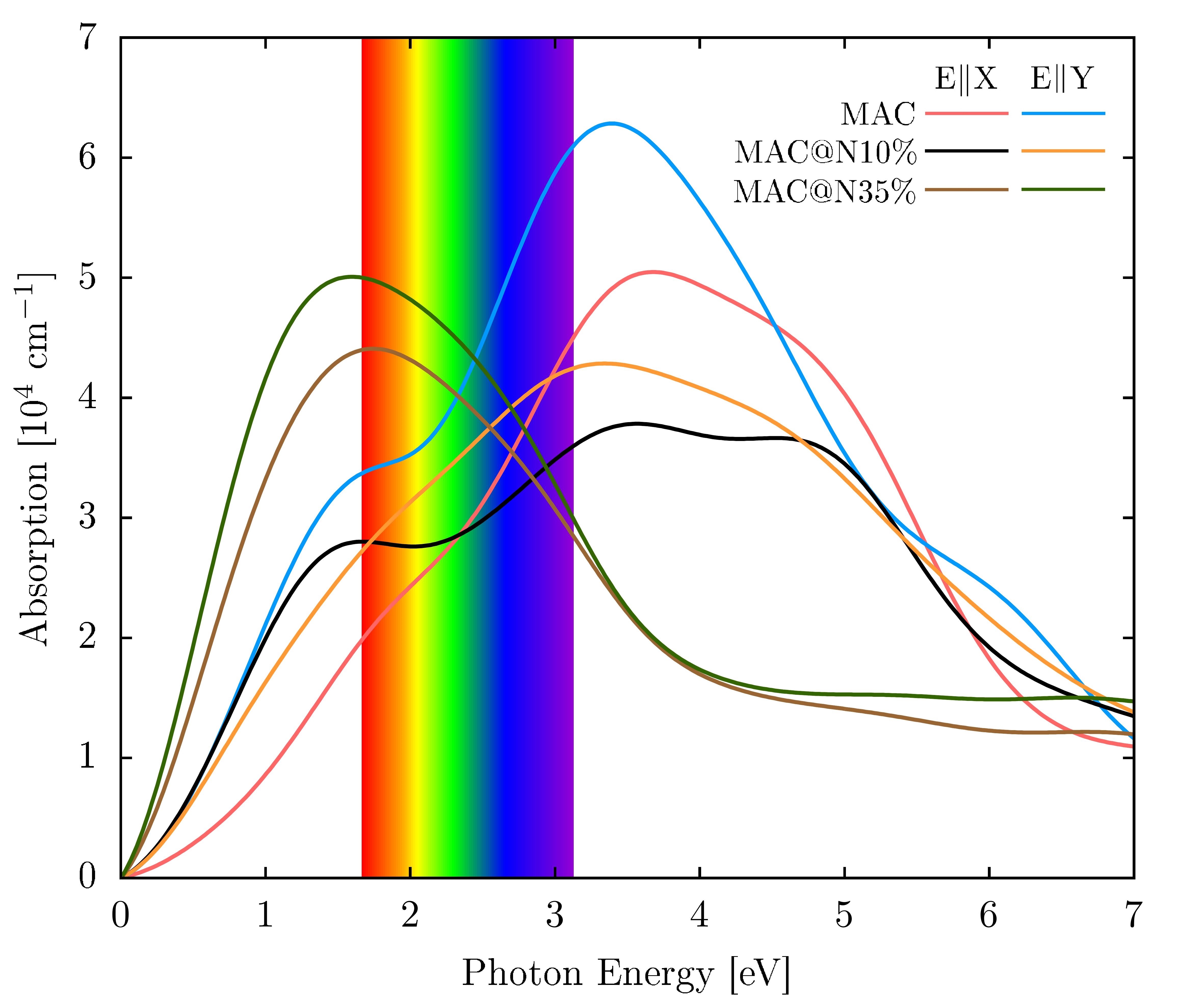}
    \caption{Optical absorption spectra of (red and blue) MAC, (black and yellow) MAC@N-10\%, and (brown and green) MAC@N-35\%. The light absorption coefficient is shown as a function of photon energy value, highlighting the influence of nitrogen doping on the optical properties.}
    \label{fig:optical}
\end{figure}

Now, we discuss the optical absorption of the MAC and MAC@N systems. In Figure \ref{fig:optical}, we present the optical absorption spectra of the undoped MAC (red and blue), MAC@N-10\% (black and yellow), and MAC@N-35\% (brown and green) for light polarization along the plane (x,y)-directions. The undoped MAC exhibits a strong optical absorption, primarily in the ultraviolet (UV) region. This behavior can be attributed to the nature of the electronic transitions in the all-carbon amorphous network. The delocalized $\pi$-electrons in the sp$^{2}$ regions of MAC contribute to transitions requiring higher photon energies, which correspond to UV light.

As observed for the MAC@N systems, introducing nitrogen into the MAC structure results in a redshift in the absorption spectra. This shift can be attributed to new localized states within the electronic bandstructure caused by nitrogen incorporation. Due to its higher electronegativity than carbon, nitrogen introduces electronic states closer to the Fermi level, lowering the energy required for electronic transitions. Consequently, lower-energy photons, corresponding to the visible or even infrared regions, become sufficient to induce these transitions, shifting the absorption peak toward longer wavelengths (lower energies). Moreover, nitrogen doping disrupts the sp$^{2}$ bonding network of the carbon atoms, increasing light scattering and reducing the intensity of the absorption peaks.

The MAC@N 10\% system exhibits an increase/decrease in optical activity in the visible/UV region (black and yellow curves), as shown in Figure \ref{fig:optical}. This moderate optical activity arises from a balance between retaining sufficient carbon-carbon $\pi$-bonding regions, which continue contributing to UV absorption, and introducing nitrogen states that induce lower-energy transitions. The 10\% nitrogen concentration is low enough to preserve some of the intrinsic electronic transitions of undoped MAC while providing enough localized states to extend the absorption into the visible spectrum.

In contrast, the MAC@N-35\% system exhibits an absorption spectrum that extends further into the infrared and visible regions (brown and green curves in Figure \ref{fig:optical}). At this higher nitrogen concentration, the additional localized states introduced by nitrogen atoms significantly reshape the material's electronic structure. The increased nitrogen content results in a higher density of states near the Fermi level, making low-energy electronic transitions more likely. As a result, the absorption spectrum is dominated by transitions involving these states, leading to optical activity that spans into the infrared and visible regions. This characteristic makes the MAC@N system with 35\% doping particularly promising for applications that require an optical response in the lower-energy range of the spectrum, such as infrared detectors and specialized optoelectronic devices.

\begin{figure}[!htb]
    \centering
    \includegraphics[width=\linewidth]{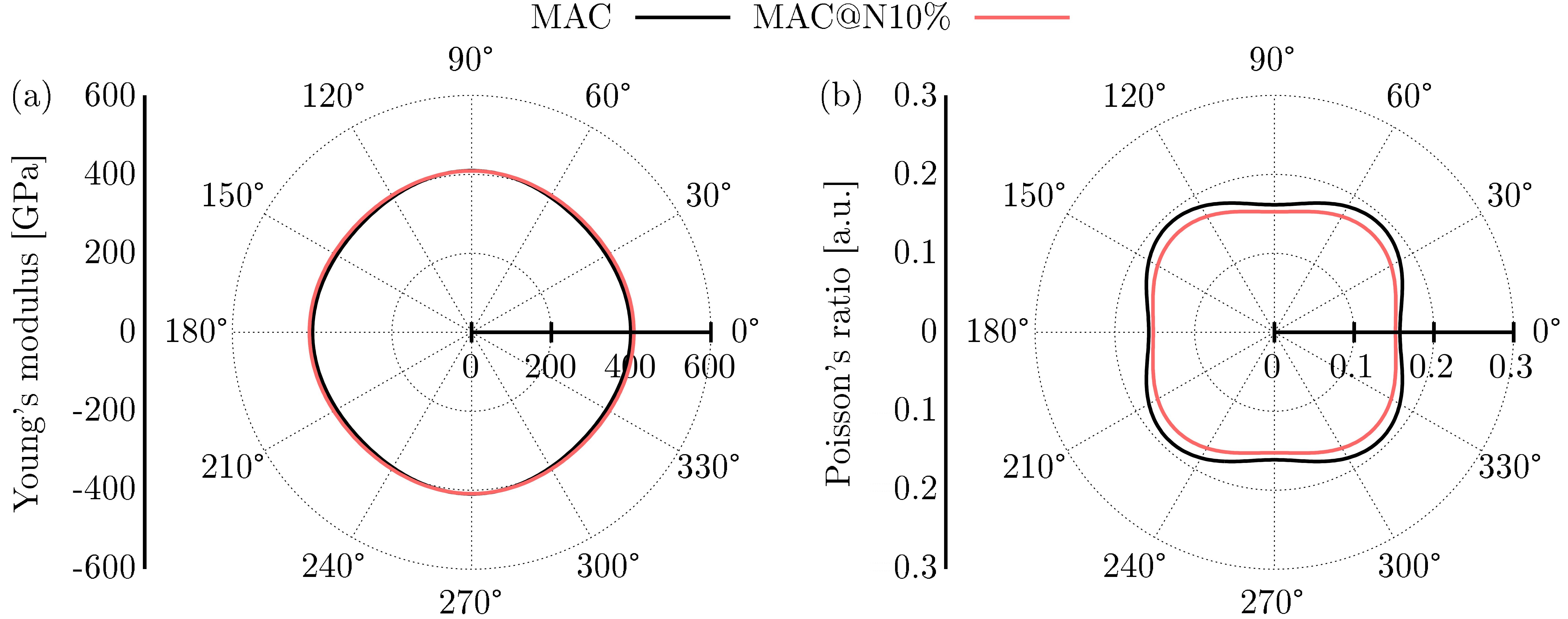}
    \caption{(a) Estimated Young's modulus and (b) Poisson's ratio values about the basal plane for MAC and MAC@N-10\%.}
    \label{fig:mechprop}
\end{figure}

Some of the mechanical properties of the MAC and MAC@N-10\% systems are illustrated in Figure \ref{fig:mechprop}. We have estimated Young's modulus ($Y(\theta)$) and  Poisson's ratio ($\nu(\theta)$) values under pressure along the xy plane, as given by the following equations \cite{doi:10.1021/acsami.9b10472,doi:10.1021/acs.jpclett.8b00616}: 

\begin{equation}
    \displaystyle Y(\theta) = \frac{{C_{11}C_{22} - C_{12}^2}}{{C_{11}\alpha^4 + C_{22}\beta^4 + \left(\frac{{C_{11}C_{22} - C_{12}^2}}{{C_{44}}} - 2.0C_{12}\right)\alpha^2\beta^2}}
    \label{young}
\end{equation}

\noindent and 

\begin{equation}
    \displaystyle \nu(\theta)= \frac{{(C_{11} + C_{22} - \frac{{C_{11}C_{22} - C_{12}^2}}{{C_{44}}})\alpha^2\beta^2 - C_{12}(\alpha^4 + \beta^4)}}{{C_{11}\alpha^4 + C_{22}\beta^4 + \left(\frac{{C_{11}C_{22} - C_{12}^2}}{{C_{44}}} - 2.0C_{12}\right)\alpha^2\beta^2}},
    \label{poisson}
\end{equation}

\noindent where, $\alpha=\cos(\theta)$ and $\beta=\sin(\theta)$. 

It is important to remark that the mechanical properties of the MAC@N system with 35\% nitrogen doping were not calculated, as this structure is only metastable at 300K, as discussed below. Due to its instability at room temperature, any mechanical property analysis would not accurately reflect the behavior of a stable lattice. Consequently, we focused our calculations on the undoped MAC and the MAC@N system with 10\% nitrogen doping, which maintains structural stability under the considered conditions.

In Figure \ref{fig:mechprop}(a), we present the results for Young's modulus for undoped MAC (black) and MAC@N-10\% (red). The estimated values are 409.40 GPa and 416.43 GPa, respectively. The variation in these values can be attributed to the impact of nitrogen atoms on the structural rigidity of the amorphous carbon network. The well-distributed sp$^{2}$ and sp$^{3}$-like hybridized carbon atoms in the undoped MAC contribute to a relatively small Young's modulus value. Introducing 10\% nitrogen slightly increases this value due to the localized disruptions caused by nitrogen atoms within the carbon lattice and nitrogen-carbon bonds. Nitrogen atoms also introduce distortions and break the uniformity of the carbon-carbon bonding network, increasing the overall stiffness of the structure. These findings agree with an MD study on undoped and nitrogen-doped graphene, in which the latter presenters a slightly higher Young's modulus than the former \cite{mortazavi2012nitrogen}. This study has also revealed that at some doping levels, up to 6\%, Young’s modulus of graphene was not changed \cite{mortazavi2012nitrogen}. 

Figure \ref{fig:mechprop}(b) illustrates the Poisson's ratio values for the same set of structures. The all-carbon MAC exhibits a slightly higher Poisson's ratio than its nitrogen-doped counterpart, and a general trend can be inferred where the Poisson's ratio decreases as the nitrogen doping concentration increases. The Poisson's ratio measures the material's tendency to expand/compress laterally when compressed/stretched longitudinally. The presence of nitrogen atoms alters the local bonding environment and introduces localized regions of stiffness, as mentioned above. This localized stiffness reduces the material's ability to deform laterally under strain, resulting in a lower Poisson's ratio.

The elastic constants, C$_{11}$, C$_{12}$, $C_{22}$, and C$_{44}$, used to calculate these mechanical properties are presented in Table \ref{tab:elastic}. These constants provide insights into the material's response to various deformations, with C$_{11}$ and C$_{22}$ reflecting the stiffness along different directions, C$_{12}$ representing the coupling between these directions, and C$_{44}$ reflecting the shear resistance. The values satisfy the Born-Huang stability criteria, i.e., ($C_{11}C_{22} - C_{12}^2>0$ and $C_{44}>0$) \cite{PhysRevB.90.224104,doi:10.1021/acs.jpcc.9b09593}, confirming that the undoped MAC and MAC@N-10\% structures maintain their structural integrity under strain. The variations in Young's modulus and Poisson's ratio with nitrogen concentration highlight the tunable nature of the mechanical properties in MAC@N systems, making them promising candidates for applications where specific mechanical characteristics are required.

\begin{table}[!htb]
\centering
\caption{Elastic constants C$_{ij}$ (GPa) and maximum values for Young's modulus (GPa) ($Y_{MAX}$) and maximum ($\nu_{MAX}$) and ($\nu_{MIN}$) Poisson's ratios.}
\label{tab:elastic}
\begin{tabular}{ l c c c c c c c c}
\hline
 Structure & C$_{11}$ & C$_{12}$ &C$_{22}$ &C$_{44}$ & $Y_{MAX}$  & $\nu_{MAX}$ & $\nu_{MIN}$ \\
 \hline
 \hline
 MAC         & $409.40$       & $66.11$     & $419.95$     & $163.56$  & $405$ & $0.18$ & $0.15$ \\
 MAC@N10\%   & $416.43$       & $63.54$     & $418.36$     & $168.94$  & $410$ & $0.17$ & $0.15$ \\
 \hline
 \hline
 \end{tabular}
\end{table}

\begin{figure}[!htb]
    \centering
    \includegraphics[width=\linewidth]{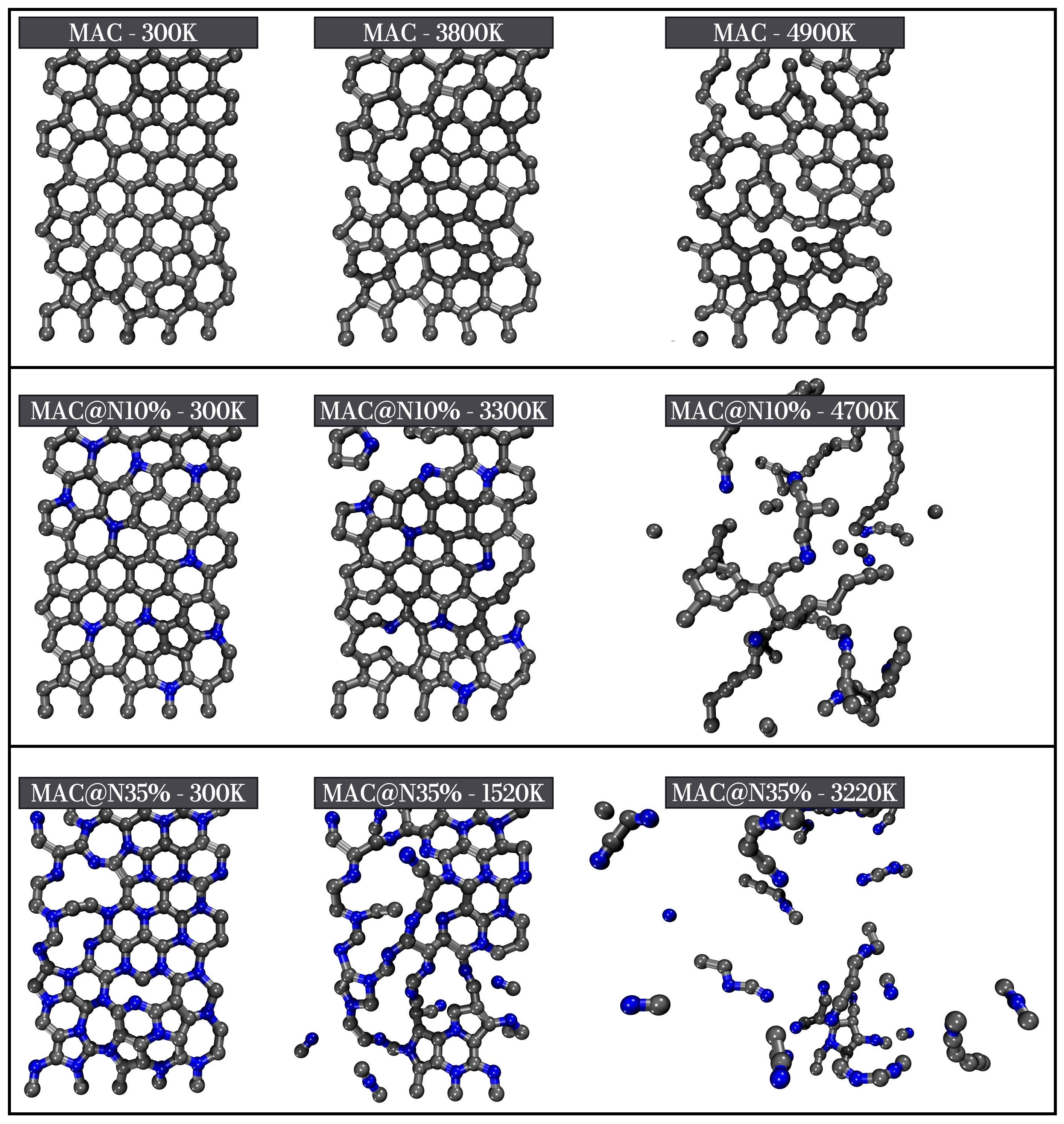}
    \caption{Representative snapshots from MD simulations illustrating the thermal stability of (top) MAC, (middle) MAC@N-10\%, and (bottom) MAC@N-35\% systems.}
    \label{fig:melting}
\end{figure}

Finally, in Figure \ref{fig:melting}, we present the results from the MD simulations, showcasing the thermal stability of the MAC and MAC@N systems, with representative snapshots taken at different temperatures. The top, middle, and bottom rows correspond to the undoped MAC, MAC@N-10\%, and MAC@N-35\% cases. These simulations provide valuable insights into these materials' structural integrity under increasing temperatures.

The undoped MAC (top row) and MAC@N-10\% (middle row) demonstrate thermal stability at room temperature, maintaining their structural integrity. The undoped MAC and MAC@N-10\% become thermally unstable at 3800K and 3300K, respectively. This difference can be attributed to nitrogen atoms in the doped system and all distinct degrees of thermal fluctuations. Incorporating nitrogen into the amorphous carbon lattice introduces a different interplay between localized states and bonding interactions, requiring lower thermal energy to disrupt the network. As a result, the nitrogen-doped MAC@N-10\% structure exhibits a slightly decreased resistance to thermal fluctuations, resulting in a lower temperature for loss of its stability. These critical temperatures for thermal stability are consistent with those reported for other MAC structures \cite{felix2020mechanical,junior2021reactive}, highlighting the inherent robustness of the amorphous carbon network.

At elevated temperatures, the MAC@N-10\% system exhibits lattice atomization and the formation of linear atomic chains (LACs) around 4700K, as seen in the sequence of snapshots illustrated in the middle row of Figure \ref{fig:melting}. This phenomenon, characterized by the appearance of elongated chains of atoms, is not observed in the undoped MAC, even at a higher temperature of 4900K. Nitrogen in the MAC@N structure facilitates the LAC appearance by forming more flexible and less constrained bonding configurations when subjected to high thermal energy values. The nitrogen atoms can create localized regions of weaker bonds, allowing the carbon network to reconfigure into LACs as the structure destabilizes.

The MAC@N-35\% case can be considered metastable at temperatures ranging from 100K to 300K. At these temperatures, some of the atomic bonds within the structure begin to reconfigure, which indicates that the lattice is under thermal strain. However, despite these reconfigurations, the atoms remain bonded, preventing the formation of LACs or atomization. This behavior suggests that while the system is not entirely stable, it maintains some structural integrity.

Additionally, within this temperature range, we can observe the presence of well-structured lattice domains where original MAC regions persist. However, the high nitrogen content introduces significant local strain and deformations, creating small holes or vacancies within the amorphous structure, as illustrated in the first snapshot of the bottom row of Figure \ref{fig:melting}. These defects contribute to the system's metastability, creating points of localized instability without leading to a complete breakdown/fracture of the overall structure. The combination of partial reconfiguration and the coexistence of stable domains with defects characterizes the metastable nature of the MAC@N-35\% system.

The rapid destabilization of MAC@N-35\% results from the high concentration of nitrogen atoms, which introduces considerable intrinsic strain and disorder into the MAC lattice, as discussed above. The increased number of nitrogen atoms disrupts the carbon-carbon bonds pattern, leading to a structure that cannot maintain its integrity even at relatively low thermal energies. As the temperature further increases, lattice atomization and the formation of LACs occur at around 1520K, lower than the temperatures required for the undoped MAC and MAC@N-10\% systems. The complete atomization of the MAC@N-35\% system occurs at 3220K.

\section{Conclusions}

In summary, we have investigated the recently synthesized nitrogen-doped amorphous monolayer carbon's structural, electronic, optical, and thermal properties using the DFTB+ approach. The results offer valuable insights into how varying nitrogen doping concentrations affect the stability and multifunctionality of MAC@N, with implications for potential applications in flexible electronics and optoelectronic devices.

Our findings reveal that the maximum nitrogen concentration for achieving a stable MAC@N structure is 35\%, beyond which the lattice becomes structurally unstable. Nitrogen doping influences the formation energy values, which are consistently higher for MAC@N than graphene@N for all concentration cases investigated here. This trend is attributed to the disordered nature of the amorphous carbon network in MAC@N, which experiences more significant strain and bond rearrangement upon nitrogen incorporation than the graphene case.

The electronic properties of MAC and MAC@N highlight nitrogen doping's effects on the electronic band structure. While the undoped MAC retains a metallic character with a Dirac-like cone, upon nitrogen doping, it disappears. The doping also creates a significant number of localized states. These changes in electronic properties suggest that nitrogen doping can be exploited to tune MAC@N's conductive behavior. 

Optical absorption spectra of MAC and MAC@N systems show that nitrogen doping causes a redshift in the absorption spectra, broadening the range of optical activity. Undoped MAC displays strong absorption, mainly in the ultraviolet range. In contrast, MAC@N has absorption into the visible and infrared regions. This shift is particularly pronounced in the MAC@N system with 35\% doping, where introducing additional nitrogen states enables lower-energy transitions. The tunable optical properties of MAC@N highlight its potential for optoelectronic applications, such as sensors and photodetectors.

As assessed through MD simulations, the thermal stability of the MAC and MAC@N systems varies significantly with nitrogen concentration. Undoped MAC and MAC@N with 10\% nitrogen doping exhibit considerably high thermal stability, with melting/vaporization temperatures of 3800K and 3300K, respectively. At 35\% nitrogen doping, MAC@N shows a significant decrease in structural stability, being metastable for temperatures ranging from 100K up to 300K, caused by increased structural disorder. The high nitrogen content destabilizes the lattice even at lower temperatures and weakens the bonding network. The estimated Young's modulus values for the undoped MAC and MACN@-10\% are about 409.4 GPa and 416.4 GPa, respectively. The variation in these values can be attributed to the impact of nitrogen atoms on the structural rigidity of the amorphous carbon network. 

\section{Acknowledgements}
This work received partial support from the Brazilian agencies CAPES, CNPq, and FAPDF.
M.L.P.J. acknowledges the financial support of the FAP-DF grant 00193-00001807/2023-16. The authors also acknowledge support from CENAPAD-SP (National High-Performance Center in São Paulo, State University of Campinas -- UNICAMP, projects: proj950, proj634 and proj960) and NACAD (High-Performance Computing Center, Lobo Carneiro Supercomputer, Federal University of Rio de Janeiro -- UFRJ, projects: a22002 and a23003) for the computational support provided. 
L.A.R.J. acknowledges the financial support from FAP-DF grants 00193.00001808/2022-71 and $00193-00001857/2023-95$, FAPDF-PRONEM grant $00193.00001247/2021-20$, and PDPG-FAPDF-CAPES Centro-Oeste grant $00193-00000867/2024-94$. L.A.R.J also acknowledges CNPq grants $350176/2022-1$ and 167745/2023-9.
D. S. G. acknowledges the Center for Computing in Engineering and Sciences at Unicamp for financial support through the FAPESP/CEPID Grant \#2013/08293-7.
%

\bibliographystyle{unsrt}
\bibliography{bibliography.bib}

\begin{thebibliography}{10}

\bibitem{hirsch2010era}
Andreas Hirsch.
\newblock The era of carbon allotropes.
\newblock {\em Nature materials}, 9(11):868--871, 2010.

\bibitem{geim2009graphene}
Andre~Konstantin Geim.
\newblock Graphene: status and prospects.
\newblock {\em science}, 324(5934):1530--1534, 2009.

\bibitem{jana2021emerging}
Susmita Jana, Arka Bandyopadhyay, Sujoy Datta, Debaprem Bhattacharya, and Debnarayan Jana.
\newblock Emerging properties of carbon based 2d material beyond graphene.
\newblock {\em Journal of Physics: Condensed Matter}, 34(5):053001, 2021.

\bibitem{kumar2018recent}
Rajesh Kumar, Ednan Joanni, Rajesh~K Singh, Dinesh~P Singh, and Stanislav~A Moshkalev.
\newblock Recent advances in the synthesis and modification of carbon-based 2d materials for application in energy conversion and storage.
\newblock {\em Progress in Energy and Combustion Science}, 67:115--157, 2018.

\bibitem{enyashin2011graphene}
Andrey~N Enyashin and Alexander~L Ivanovskii.
\newblock Graphene allotropes.
\newblock {\em physica status solidi (b)}, 248(8):1879--1883, 2011.

\bibitem{fan2021biphenylene}
Qitang Fan, Linghao Yan, Matthias~W Tripp, Ond{\v{r}}ej Krej{\v{c}}{\'\i}, Stavrina Dimosthenous, Stefan~R Kachel, Mengyi Chen, Adam~S Foster, Ulrich Koert, Peter Liljeroth, et~al.
\newblock Biphenylene network: A nonbenzenoid carbon allotrope.
\newblock {\em Science}, 372(6544):852--856, 2021.

\bibitem{hou2022synthesis}
Lingxiang Hou, Xueping Cui, Bo~Guan, Shaozhi Wang, Ruian Li, Yunqi Liu, Daoben Zhu, and Jian Zheng.
\newblock Synthesis of a monolayer fullerene network.
\newblock {\em Nature}, 606(7914):507--510, 2022.

\bibitem{meirzadeh2023few}
Elena Meirzadeh, Austin~M Evans, Mehdi Rezaee, Milena Milich, Connor~J Dionne, Thomas~P Darlington, Si~Tong Bao, Amymarie~K Bartholomew, Taketo Handa, Daniel~J Rizzo, et~al.
\newblock A few-layer covalent network of fullerenes.
\newblock {\em Nature}, 613(7942):71--76, 2023.

\bibitem{toh2020synthesis}
Chee-Tat Toh, Hongji Zhang, Junhao Lin, Alexander~S Mayorov, Yun-Peng Wang, Carlo~M Orofeo, Darim~Badur Ferry, Henrik Andersen, Nurbek Kakenov, Zenglong Guo, et~al.
\newblock Synthesis and properties of free-standing monolayer amorphous carbon.
\newblock {\em Nature}, 577(7789):199--203, 2020.

\bibitem{tian2023disorder}
Huifeng Tian, Yinhang Ma, Zhenjiang Li, Mouyang Cheng, Shoucong Ning, Erxun Han, Mingquan Xu, Peng-Fei Zhang, Kexiang Zhao, Ruijie Li, et~al.
\newblock Disorder-tuned conductivity in amorphous monolayer carbon.
\newblock {\em Nature}, 615(7950):56--61, 2023.

\bibitem{bai2024nitrogen}
Xiuhui Bai, Pengfei Hu, Ang Li, Youwei Zhang, Aowen Li, Guangjie Zhang, Yufeng Xue, Tianxing Jiang, Zezhou Wang, Hanke Cui, et~al.
\newblock Nitrogen-doped amorphous monolayer carbon.
\newblock {\em Nature}, pages 1--5, 2024.

\bibitem{zhang2022structure}
Yu-Tian Zhang, Yun-Peng Wang, Xianli Zhang, Yu-Yang Zhang, Shixuan Du, and Sokrates~T Pantelides.
\newblock Structure of amorphous two-dimensional materials: elemental monolayer amorphous carbon versus binary monolayer amorphous boron nitride.
\newblock {\em Nano Letters}, 22(19):8018--8024, 2022.

\bibitem{joo2017realization}
Won-Jae Joo, Jae-Hyun Lee, Yamujin Jang, Seog-Gyun Kang, Young-Nam Kwon, Jaegwan Chung, Sangyeob Lee, Changhyun Kim, Tae-Hoon Kim, Cheol-Woong Yang, et~al.
\newblock Realization of continuous zachariasen carbon monolayer.
\newblock {\em Science advances}, 3(2):e1601821, 2017.

\bibitem{felix2020mechanical}
Levi~C Felix, Raphael~M Tromer, Pedro~AS Autreto, Luiz~A Ribeiro~Junior, and Douglas~S Galvao.
\newblock On the mechanical properties and thermal stability of a recently synthesized monolayer amorphous carbon.
\newblock {\em The Journal of Physical Chemistry C}, 124(27):14855--14860, 2020.

\bibitem{junior2021reactive}
Marcelo Lopes~Pereira Junior, Wiliam~Ferreira da~Cunha, Douglas~Soares Galv{\~a}o, and Luiz Antonio~Ribeiro Junior.
\newblock A reactive molecular dynamics study on the mechanical properties of a recently synthesized amorphous carbon monolayer converted into a nanotube/nanoscroll.
\newblock {\em Physical chemistry chemical physics}, 23(15):9089--9095, 2021.

\bibitem{gastellu2022electronic}
Nicolas Gastellu, Michael Kilgour, and Lena Simine.
\newblock Electronic conduction through monolayer amorphous carbon nanojunctions.
\newblock {\em The Journal of Physical Chemistry Letters}, 13(1):339--344, 2022.

\bibitem{xia2020monolayer}
Rui Xia, Songbo Chen, Subin Jiang, Jingyan Zhang, Xing Wang, Changqi Sun, Yongcheng Xiao, Yonggang Liu, and Meizhen Gao.
\newblock Monolayer amorphous carbon-bridged nanosheet mesocrystal: Facile preparation, morphosynthetic transformation, and energy storage applications.
\newblock {\em ACS Applied Materials \& Interfaces}, 13(1):1114--1126, 2020.

\bibitem{garzon2022optoelectronic}
Antonio~J Garz{\'o}n-Ram{\'\i}rez, Nicolas Gastellu, and Lena Simine.
\newblock Optoelectronic current through unbiased monolayer amorphous carbon nanojunctions.
\newblock {\em The Journal of Physical Chemistry Letters}, 13(4):1057--1062, 2022.

\bibitem{avouris2012graphene}
Phaedon Avouris and Christos Dimitrakopoulos.
\newblock Graphene: synthesis and applications.
\newblock {\em Materials today}, 15(3):86--97, 2012.

\bibitem{mortazavi2019prediction}
Bohayra Mortazavi, Masoud Shahrokhi, Alexander~V Shapeev, Timon Rabczuk, and Xiaoying Zhuang.
\newblock Prediction of c 7 n 6 and c 9 n 4: stable and strong porous carbon-nitride nanosheets with attractive electronic and optical properties.
\newblock {\em Journal of Materials Chemistry C}, 7(35):10908--10917, 2019.

\bibitem{mortazavi2019outstanding}
Bohayra Mortazavi, Masoud Shahrokhi, Mostafa Raeisi, Xiaoying Zhuang, Luiz Felipe~C Pereira, and Timon Rabczuk.
\newblock Outstanding strength, optical characteristics and thermal conductivity of graphene-like bc3 and bc6n semiconductors.
\newblock {\em Carbon}, 149:733--742, 2019.

\bibitem{mortazavi2018n}
Bohayra Mortazavi, Meysam Makaremi, Masoud Shahrokhi, Zheyong Fan, and Timon Rabczuk.
\newblock N-graphdiyne two-dimensional nanomaterials: Semiconductors with low thermal conductivity and high stretchability.
\newblock {\em Carbon}, 137:57--67, 2018.

\bibitem{mortazavi2023electronic}
Bohayra Mortazavi.
\newblock Electronic, thermal and mechanical properties of carbon and boron nitride holey graphyne monolayers.
\newblock {\em Materials}, 16(20):6642, 2023.

\bibitem{mortazavi2020nanoporous}
Bohayra Mortazavi, Fazel Shojaei, Masoud Shahrokhi, Maryam Azizi, Timon Rabczuk, Alexander~V Shapeev, and Xiaoying Zhuang.
\newblock Nanoporous c3n4, c3n5 and c3n6 nanosheets; novel strong semiconductors with low thermal conductivities and appealing optical/electronic properties.
\newblock {\em Carbon}, 167:40--50, 2020.

\bibitem{mortazavi2021ultrahigh}
Bohayra Mortazavi.
\newblock Ultrahigh thermal conductivity and strength in direct-gap semiconducting graphene-like bc6n: A first-principles and classical investigation.
\newblock {\em Carbon}, 182:373--383, 2021.

\bibitem{mortazavi2022anisotropic}
Bohayra Mortazavi, Fazel Shojaei, Mehmet Yagmurcukardes, Alexander~V Shapeev, and Xiaoying Zhuang.
\newblock Anisotropic and outstanding mechanical, thermal conduction, optical, and piezoelectric responses in a novel semiconducting bcn monolayer confirmed by first-principles and machine learning.
\newblock {\em Carbon}, 200:500--509, 2022.

\bibitem{wu2016tuning}
Jingjie Wu, Marco-Tulio~F Rodrigues, Robert Vajtai, and Pulickel~M Ajayan.
\newblock Tuning the electrochemical reactivity of boron-and nitrogen-substituted graphene.
\newblock {\em Advanced Materials}, 28(29):6239--6246, 2016.

\bibitem{talukder2021nitrogen}
Niladri Talukder, Yudong Wang, Bharath~Babu Nunna, and Eon~Soo Lee.
\newblock Nitrogen-doped graphene nanomaterials for electrochemical catalysis/reactions: A review on chemical structures and stability.
\newblock {\em Carbon}, 185:198--214, 2021.

\bibitem{hourahine2020dftb+}
Ben Hourahine, B{\'a}lint Aradi, Volker Blum, Frank Bonafe, Alex Buccheri, Cristopher Camacho, Caterina Cevallos, MY~Deshaye, T~Dumitric{\u{a}}, A~Dominguez, et~al.
\newblock Dftb+, a software package for efficient approximate density functional theory based atomistic simulations.
\newblock {\em The Journal of chemical physics}, 152(12), 2020.

\bibitem{aradi2007dftb+}
Balint Aradi, Ben Hourahine, and Th~Frauenheim.
\newblock Dftb+, a sparse matrix-based implementation of the dftb method.
\newblock {\em The Journal of Physical Chemistry A}, 111(26):5678--5684, 2007.

\bibitem{gaus2013parametrization}
Michael Gaus, Albrecht Goez, and Marcus Elstner.
\newblock Parametrization and benchmark of dftb3 for organic molecules.
\newblock {\em Journal of Chemical Theory and Computation}, 9(1):338--354, 2013.

\bibitem{gaus2014parameterization}
Michael Gaus, Xiya Lu, Marcus Elstner, and Qiang Cui.
\newblock Parameterization of dftb3/3ob for sulfur and phosphorus for chemical and biological applications.
\newblock {\em Journal of chemical theory and computation}, 10(4):1518--1537, 2014.

\bibitem{shuichi1991constant}
Nos{\'e} Shuichi.
\newblock Constant temperature molecular dynamics methods.
\newblock {\em Progress of Theoretical Physics Supplement}, 103:1--46, 1991.

\bibitem{shi2015much}
Zhiming Shi, Alex Kutana, and Boris~I Yakobson.
\newblock How much n-doping can graphene sustain?
\newblock {\em The Journal of Physical Chemistry Letters}, 6(1):106--112, 2015.

\bibitem{tromer2024structural}
Raphael~M Tromer, Marcelo L~Pereira J{\'u}nior, Luiz A~Ribeiro J{\'u}nior, and Douglas~S Galv{\~a}o.
\newblock Structural and electronic properties of amorphous silicon and germanium monolayers and nanotubes: A dft investigation.
\newblock {\em Chemical Physics Letters}, page 141647, 2024.

\bibitem{tromer2021optoelectronic}
Raphael~M Tromer, Levi~C Felix, Luiz~A Ribeiro, and Douglas~S Galvao.
\newblock Optoelectronic properties of amorphous carbon-based nanotube and nanoscroll.
\newblock {\em Physica E: Low-dimensional Systems and Nanostructures}, 130:114683, 2021.

\bibitem{doi:10.1021/acsami.9b10472}
Bing Wang, Qisheng Wu, Yehui Zhang, Liang Ma, and Jinlan Wang.
\newblock Auxetic b4n monolayer: A promising 2d material with in-plane negative poisson’s ratio and large anisotropic mechanics.
\newblock {\em ACS Applied Materials \& Interfaces}, 11(36):33231--33237, 2019.

\bibitem{doi:10.1021/acs.jpclett.8b00616}
Yu~Zhao, Xiaoyin Li, Junyi Liu, Cunzhi Zhang, and Qian Wang.
\newblock A new anisotropic dirac cone material: A b2s honeycomb monolayer.
\newblock {\em The Journal of Physical Chemistry Letters}, 9(7):1815--1820, 2018.

\bibitem{mortazavi2012nitrogen}
Bohayra Mortazavi, Said Ahzi, Val{\'e}rie Toniazzo, and Yves R{\'e}mond.
\newblock Nitrogen doping and vacancy effects on the mechanical properties of graphene: A molecular dynamics study.
\newblock {\em Physics Letters A}, 376(12-13):1146--1153, 2012.

\bibitem{PhysRevB.90.224104}
F\'elix Mouhat and Fran\ifmmode \mbox{\c{c}}\else \c{c}\fi{}ois-Xavier Coudert.
\newblock Necessary and sufficient elastic stability conditions in various crystal systems.
\newblock {\em Phys. Rev. B}, 90:224104, Dec 2014.

\bibitem{doi:10.1021/acs.jpcc.9b09593}
Yiran Ying, Ke~Fan, Sicong Zhu, Xin Luo, and Haitao Huang.
\newblock Theoretical investigation of monolayer rhtecl semiconductors as photocatalysts for water splitting.
\newblock {\em The Journal of Physical Chemistry C}, 124(1):639--646, 2020.

\end{thebibliography}
\end{document}